\documentclass[jkps,preprint,fleqn,showpacs,showkeys]{revtex4}
\usepackage{graphicx}
\usepackage{amssymb}
\usepackage{amsmath}
\usepackage{bm}
\usepackage{textcomp}
\begin{document}
\setcounter{page}{0}
\title[]{Meshfree Local Radial Basis Function Collocation Method with
Image Nodes}
\author{Seung Ki \surname{Baek}}
\email{seungki@pknu.ac.kr}
\thanks{Fax: +82-51-629-5549}
\affiliation{Department of Physics, Pukyong National University, Busan 48513,
Korea}
\author{Minjae \surname{Kim}}
\affiliation{Department of Physics, Pukyong National University, Busan 48513,
Korea}

\date[]{Received 6 April 2017}

\begin{abstract}
We numerically solve two-dimensional heat diffusion problems by using a simple
variant of the meshfree local radial-basis function (RBF) collocation method.
The main idea is to include an additional set of sample nodes outside the
problem domain, similarly to the method of images in electrostatics, to perform
collocation on the domain boundaries. We can thereby take into account the
temperature profile as well as its gradients specified by boundary conditions at
the same time, which holds true even for a node where two or more boundaries
meet with different boundary conditions. We argue that the image method is
computationally efficient when combined with the local RBF collocation method,
whereas the addition of image nodes becomes very costly in case of the global
collocation. We apply our modified method to a benchmark test of a boundary
value problem, and find that this simple modification reduces the maximum error
from the analytic solution significantly. The reduction is small for an initial
value problem with simpler boundary conditions. We observe increased numerical
instability, which has to be compensated for by a sufficient number of sample
nodes and/or more careful parameter choices for time integration.
\end{abstract}


\maketitle

\section{INTRODUCTION}

Numerical methods to solve a partial differential equation (PDE) are of immense
importance in various branches of science and engineering, including heat
transfer, structural mechanics, fluid mechanics, electromagnetism, quantum
mechanics, finances, and so on. The finite-difference method (FDM) is one of the
easiest to implement, but applicable to problems with relatively simple
geometry. The finite-element method (FEM) allows more flexible geometry and has
thus become the most widely used technique for many engineering applications. A
variety of FEM packages, either commercial or non-commercial, are currently
available, and they have proved the importance of numerical analysis in
industries, because the method has boosted productivity
by helping test prototype designs accurately.

The above methods need to decompose the problem domain into a mesh and use
information of neighbors on the mesh to calculate derivatives at each given
node. The construction of a mesh is often time-consuming, especially for
high-dimensional complex-shaped boundary problems, and the use
of the mesh becomes problematic when the object being simulated is deformed
largely enough to change the connectivity between neighbors. Although we may
create a new mesh during runtime, we have to assign reasonable interpolation
results to the new mesh nodes based on the existing ones, which could be
an additional source of error.
For this reason, researchers have also devised meshfree methods, which do not
require fixed connectivity between nodes.
A well-known example is the Kansa method~\cite{Kansa1990,Fasshauer2007}, which
makes use of radial basis functions (RBF) to approximate the solution of a given
PDE. This method has been successfully applied to many different
problems~\cite{Hon1998,Hon1999,Larsson2003,Perko2001,Chen2010}.

One difficulty with the Kansa method is that it is not readily scalable,
because one has to solve a linear system described by a fully populated $N
\times N$ matrix, where $N$ is the number of sample nodes in the domain of a
given PDE. The number of operations required by a direct linear solver will be
of $O(N^3)$. This is the reason that a local version of the Kansa method has
been proposed in Ref.~\citealp{Sarler2006}, because its number of required
operations would then scale linearly with $N$. The details of the method will be
given in the next section.

In this work, we show that the numerical performance of the local RBF
collocation method can be improved further by a small modification, which takes
into account the outside of the given domain, similarly to the method of images
in electrostatics~\cite{Jackson1999}.
In fact, the idea of using extra nodes outside the domain
has already been suggested by Kansa himself in Ref.~\citealp{Fedoseyev2002},
and it is called `PDE collocation on the boundary (PDECB)'.
Our point is that adding extra nodes can be very costly if the complexity grows
as $O(N^3)$.
In the local version, on the other hand, the increment of computation would be
determined by the surface-volume ratio of the system, which usually becomes
negligible when we deal with a large number of sample nodes.
In Sec.~\ref{sec:image}, we explain our results, and compare how the
results change by solving two benchmark test problems in Sec.~\ref{sec:bench}.
We then conclude this work in Sec.~\ref{sec:conclusion}.

\section{LOCAL RBF COLLOCATION METHOD BY \u{S}ARLER AND VERTNIK}

In this section, we will explain a local version of Kansa's method
in Ref.~\citealp{Sarler2006}. To illustrate the method, the authors of
Ref.~\citealp{Sarler2006} have dealt with a diffusion equation
\begin{equation}
\rho c \frac{\partial}{\partial t}T = \nabla \cdot (k \nabla T),
\label{eq:diff}
\end{equation}
where $\rho$, $c$, $t$, $T$, and $k$ denote mass density, heat capacity, time,
temperature, and thermal conductivity, respectively. The problem is defined on a
spatial domain $\Omega$ with a boundary $\Gamma$. We consider three boundary
conditions: Suppose a node on $\Gamma$, located at $\mathbf{r}$. The outward
unit normal vector on the boundary is denoted as $\mathbf{n}$. First, the
Dirichlet boundary condition fixes $T(\mathbf{r})$ to a certain value $T_D$.
Second, the Neumann boundary condition requires that the normal derivative of
$T$ should vanish so that $\nabla T \cdot \mathbf{n} = 0$. Last, the Robin
boundary condition is defined as follows:
\begin{equation}
\nabla T \cdot \mathbf{n} = R \left( T - T_{\rm ref} \right),
\label{eq:robin}
\end{equation}
where $R$ is a constant and $T_{\rm ref}$ is a reference temperature to be
prescribed by the problem.

In Ref.~\citealp{Sarler2006}, the numerical procedure to solve this PDE goes as
follows:
\begin{enumerate}
\item Sample $N_\Omega$ nodes inside $\Omega$ and $N_\Gamma$ nodes on $\Gamma$.
In total, we have $N = N_\Omega + N_\Gamma$ nodes. We have chosen a regular grid
for sampling the nodes to compare the results clearly, but the method works with
an irregular node arrangements as well.
\item For each sample node $l$, determine its domain of influence $_l \omega$.
We will focus on this particular node and its domain throughout
this explanation. Let us thus drop the index $l$ for brevity henceforth.
If $l$ lies inside $\Omega$, $\omega$ is composed of
the $K$ nearest neighbors of $l$, including $l$ itself.
See Fig.~\ref{fig:influence} with $K=5$ as an example.
If $l$ lies on $\Gamma$, it needs some care, as will be explained at the end of
this section.
\begin{figure}
\includegraphics[width=0.3\textwidth]{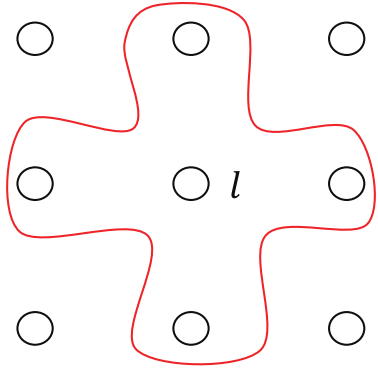}
\caption{Example of the domain of influence for node $l$ with $K=5$
nodes, all of which lie inside $\Omega$.}
\label{fig:influence}
\end{figure}
Let us denote their positions as $\mathbf{r}_n$ with $n=1, 2, \ldots,$$~K$.
Without loss of generality, we may assign $n=1$ to the focal node $l$.
\item Calculate the distance between every pair of sample nodes inside $\omega$
and define $d_0$ as the longest one. This parameter is used in the RBF for this
$\omega$, defined in a multiquadric form
\begin{equation}
\psi_k (\mathbf{r}) = \left[ d_k^2 (\mathbf{r}) + c^2 d_0^2
\label{eq:mq}
\right]^{1/2},
\end{equation}
where $d_k$ is the distance from $\mathbf{r}$ to node $k$ inside $\omega$
($k=1, 2, \ldots,$$~K$) and $c$ is a shape parameter.
\item If $l$ lies inside $\Omega$ and not on $\Gamma$, determine the
collocation coefficients $\alpha_k$'s such that reproduce the values of $T$
for all the $K$ sample nodes inside $\omega$. Specifically, we have to
solve the following set of linear equations
\begin{equation}
T( \mathbf{r}_n ) = \sum_{k=1}^{K} \psi_k (\mathbf{r}_n) \alpha_k
\label{eq:collocation}
\end{equation}
with $n=1, 2, \ldots,$$~K$. For example, if $K=5$, the equation is written
as
\begin{equation}
\left( \begin{array}{c}
T( \mathbf{r}_1 )\\
T( \mathbf{r}_2 )\\
\vdots\\
\vdots\\
T( \mathbf{r}_5 )
\end{array} \right)
=
\left( \begin{array}{cccc}
\psi_{11} & \psi_{12} & \cdots & \psi_{15}\\
\psi_{21} & \psi_{22} & \cdots & \psi_{25}\\
\vdots & \vdots & \cdots & \vdots \\
\vdots & \vdots & \ddots & \vdots \\
\psi_{51} & \psi_{52} & \cdots & \psi_{55}\\
\end{array} \right)
\left( \begin{array}{c}
\alpha_1\\
\alpha_2\\
\vdots\\
\vdots\\
\alpha_5
\end{array} \right),
\label{eq:mat1}
\end{equation}
where $\psi_{nk} \equiv \psi_k (\mathbf{r}_n)$.
Note that Eq.~(\ref{eq:mq}) is readily
differentiable so that we can approximate the derivatives of $T$ in the target
PDE by taking derivatives on the right-hand side of Eq.~(\ref{eq:collocation})
once $\alpha_k$'s are identified. By applying an explicit time integration
scheme to Eq.~(\ref{eq:diff}), calculate a new value of $T$ at the focal node
$l$. Repeat this procedure for all the $N_\Omega$ sample nodes inside $\Omega$,
and update $T$ there.
\item Now we come to the other case that $l$ lies on $\Gamma$.
Inside its domain of influence $\omega$, we may generally assume
that $K_\Omega$ nodes are domain nodes whereas the other $K_\Gamma$
nodes lie on boundaries, with $K = K_\Omega + K_\Gamma$.
In constructing a matrix equation such as
Eq.~(\ref{eq:mat1}), we use the information on the boundary conditions for the
latter $K_\Gamma$ nodes.
For example, suppose $K_\Gamma = 2$: We have $\partial T/\partial x = 0$
at $\mathbf{r}_1$ due to the Neumann boundary condition, and
the temperature is fixed to $T_D$ by the Dirichlet boundary condition at
$\mathbf{r}_2$. We thus obtain the following matrix equation
\begin{equation}
\left( \begin{array}{c}
0 \\
T_D \\
T( \mathbf{r}_3 )\\
T( \mathbf{r}_4 )\\
T( \mathbf{r}_5 )
\end{array} \right)
=
\left( \begin{array}{cccc}
\frac{\partial}{\partial x} \psi_{11} &
\frac{\partial}{\partial x} \psi_{12} &
\cdots &
\frac{\partial}{\partial x} \psi_{15}\\
\psi_{21} & \psi_{22} & \cdots & \psi_{25}\\
\psi_{31} & \psi_{32} & \cdots & \psi_{35}\\
\psi_{41} & \psi_{42} & \cdots & \psi_{45}\\
\psi_{51} & \psi_{52} & \cdots & \psi_{55}\\
\end{array} \right)
\left( \begin{array}{c}
\alpha_1\\
\alpha_2\\
\vdots\\
\vdots\\
\alpha_5
\end{array} \right).
\label{eq:mat2}
\end{equation}
Note that $T(\mathbf{r}_1)$ of the focal node $l$ is not
taken into account in determining the collocation coefficients $\alpha_k$'s,
because only the derivative of $T$ is specified by the
boundary condition. The temperature of $l$ should be updated by calculating
\begin{equation}
T( \mathbf{r}_1 ) = \sum_{k=1}^5 \psi_k ( \mathbf{r}_1) \alpha_k,
\end{equation}
after solving Eq.~(\ref{eq:mat2}) for $\alpha_k$'s. Repeat this procedure for
all the $N_\Gamma$ sample nodes on $\Gamma$.
\item Go back to Step 4 for the next time step.
\end{enumerate}

\begin{figure}
\includegraphics[width=0.3\textwidth]{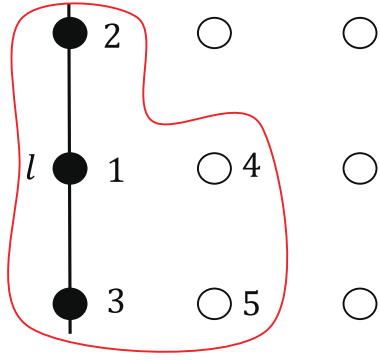}
\caption{Example of $\omega$ when $l$ lies on $\Gamma$ represented by the
vertical line. The empty and filled circles represent nodes in $\Omega$ and
those in $\Gamma$, respectively.}
\label{fig:caveat}
\end{figure}

\begin{figure}
\includegraphics[width=0.3\textwidth]{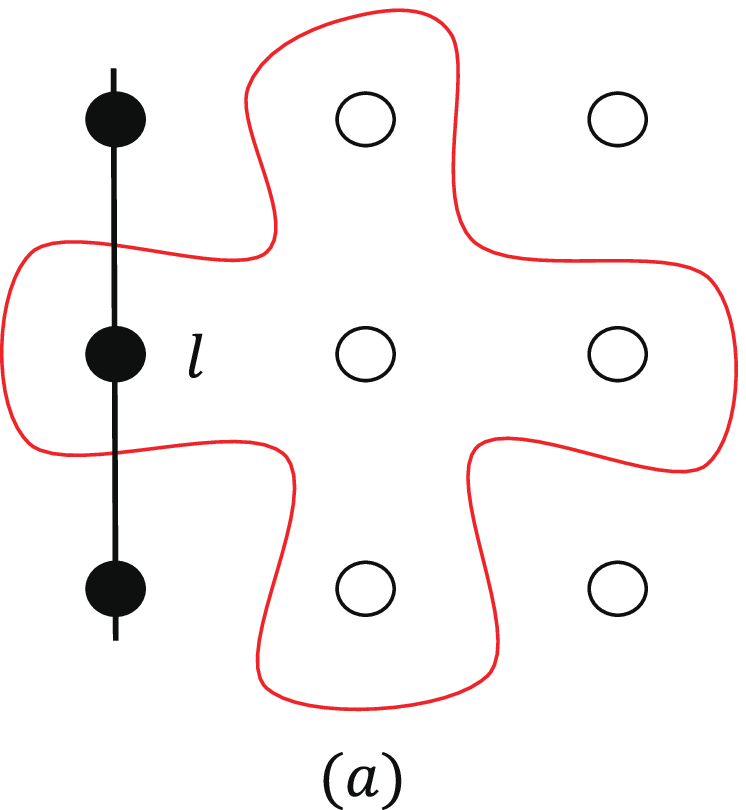}
\includegraphics[width=0.3\textwidth]{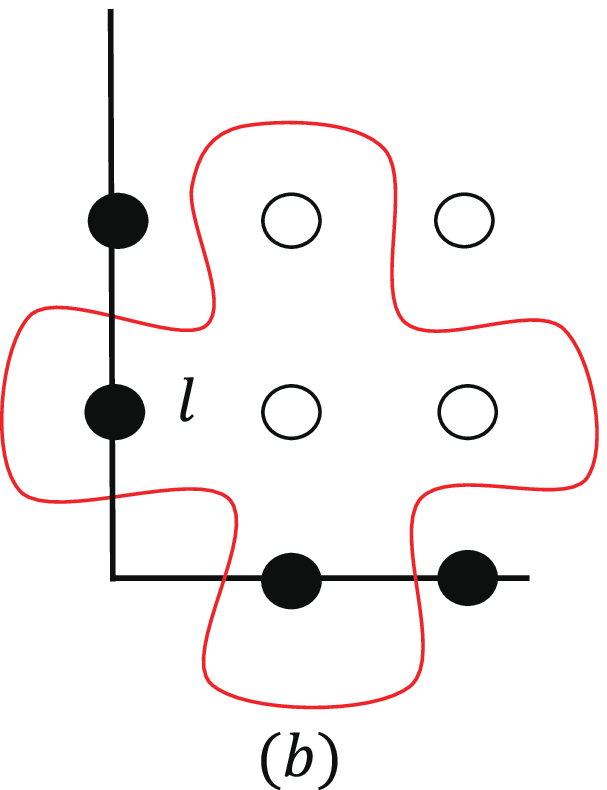}
\caption{Examples of $\omega$ to make the collocation matrix non-singular.
The lines are boundaries, and the empty and filled circles represent nodes in
$\Omega$ and those in $\Gamma$, respectively. (a) The focal node $l$ is the only
one on the boundary inside $\omega$. (b) The domain of influence $\omega$
is located on a corner and contains another boundary node than $l$.
}
\label{fig:hon}
\end{figure}

As mentioned in Step 2,
one should be careful in determining $\omega$ if the focal node $l$ belongs
to $\Gamma$. In Fig.~\ref{fig:caveat}, we construct $\omega$ by choosing the
$K=5$ nearest neighbors of $l$. Suppose that we impose the Neumann boundary
condition on this $\Gamma$. The matrix equation to solve is obtained as
\begin{equation}
\left( \begin{array}{c}
0 \\
0 \\
0 \\
T( \mathbf{r}_4 )\\
T( \mathbf{r}_5 )
\end{array} \right)
=
\left( \begin{array}{cccc}
\frac{\partial}{\partial x} \psi_{11} &
\frac{\partial}{\partial x} \psi_{12} &
\cdots &
\frac{\partial}{\partial x} \psi_{15}\\
\frac{\partial}{\partial x} \psi_{21} &
\frac{\partial}{\partial x} \psi_{22} &
\cdots &
\frac{\partial}{\partial x} \psi_{25}\\
\frac{\partial}{\partial x} \psi_{31} &
\frac{\partial}{\partial x} \psi_{32} &
\cdots &
\frac{\partial}{\partial x} \psi_{35}\\
\psi_{41} & \psi_{42} & \cdots & \psi_{45}\\
\psi_{51} & \psi_{52} & \cdots & \psi_{55}\\
\end{array} \right)
\left( \begin{array}{c}
\alpha_1\\
\alpha_2\\
\alpha_3\\
\alpha_4\\
\alpha_5
\end{array} \right).
\label{eq:mat3}
\end{equation}
Note that $\frac{\partial}{\partial x} \psi_{nk}$ is identically zero for
$1 \le n \le 3$ and $1 \le k \le 3$, because each $\psi_{nk}$ is a RBF.
Therefore, we
have only two degrees of freedom, $\alpha_4$ and $\alpha_5$, to make three
different derivatives vanish. In other words, the matrix is singular. The
problem can be avoided by defining $\omega$ in a different way so that the
number of domain nodes is greater than or equal to that of boundary nodes as
shown in Figs.~\ref{fig:hon}(a) and \ref{fig:hon}(b) (See, e.g.,
Ref.~\citealp{Hon2015}).

\section{METHOD OF IMAGES}
\label{sec:image}

The method in the previous section treats a node differently depending on
whether it belongs to $\Omega$ or $\Gamma$. That is, the present value of $T$
on the node does not appear in the collocation matrix when it is subject to a
boundary condition specified by the derivative of $T$. The reason is that one
has $K$ unknowns, which implies that the number of equations cannot be
greater than $K$,
whereas the node on $\Gamma$ introduces two equations, one for $T$ and the other
for its derivative. The situation could be worse if the node was on a corner so
that it should satisfy two or more boundary conditions at the same time.

\begin{figure}
\includegraphics[width=0.3\textwidth]{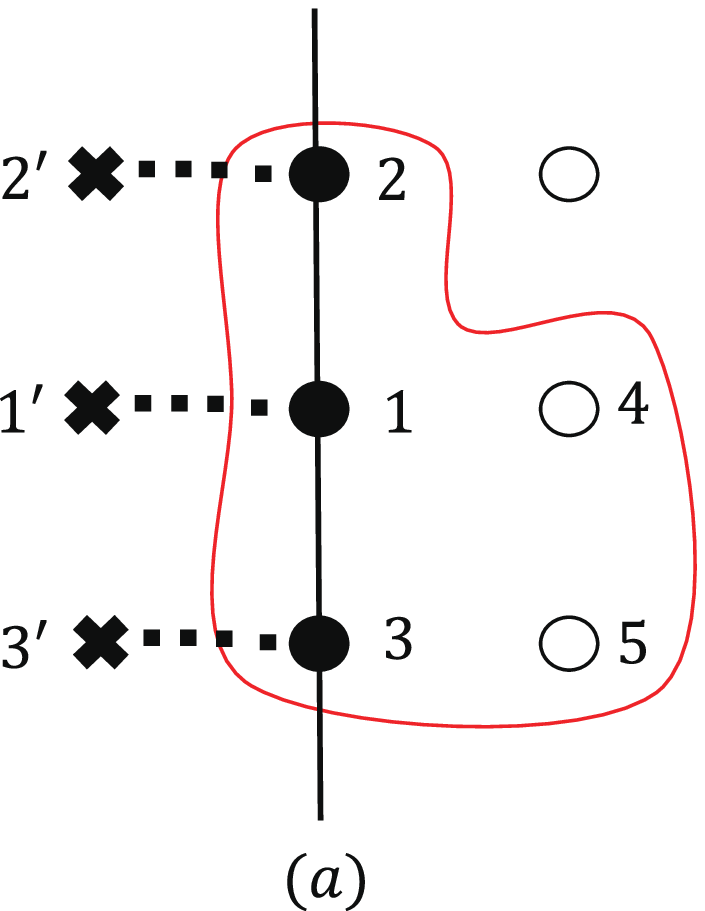}
\includegraphics[width=0.3\textwidth]{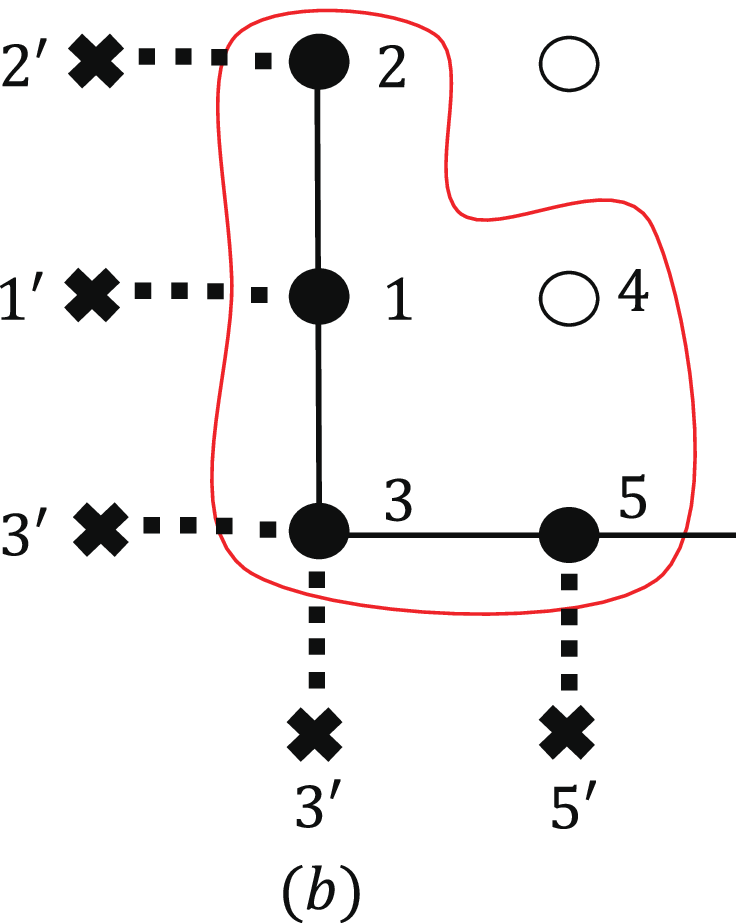}
\caption{Examples of $\omega$ with image nodes, represented by the crosses
and primed indices.
The lines are boundaries, and the empty and filled circles represent nodes in
$\Omega$ and those in $\Gamma$, respectively.
}
\label{fig:image}
\end{figure}

When we solve the Laplace equation in electrostatics,
the boundary conditions can be handled by the method of
images~\cite{Jackson1999}. Numerically, the images can be simulated
by introducing extra
nodes outside $\Omega$: They provide more unknowns, but we do not have to
consider neither $T$ nor its derivative on these nodes. The method would work
only approximately, because the RBF in Eq.~(\ref{eq:mq}) is not an
exact solution for Eq.~(\ref{eq:diff}).
Once again, there is no reason to assume such a regular grid for the image nodes
as in Fig.~\ref{fig:image}. They do not even have to be put outside
$\Omega$, as long as the collocation matrix is non-singular.

To illustrate how our method works, Fig.~\ref{fig:image}(a) shows the domain of
influence which led to a singular collocation matrix in the previous section.
This time, however, every boundary node $i$ is accompanied by an image $i'$
($i=1,2,3$).  Under the Neumann boundary condition, the collocation equation is
written as
\begin{equation}
\left( \begin{array}{c}
T( \mathbf{r}_1 )\\
T( \mathbf{r}_2 )\\
T( \mathbf{r}_3 )\\
T( \mathbf{r}_4 )\\
T( \mathbf{r}_5 )\\
0 \\
0 \\
0
\end{array} \right)
=
\left( \begin{array}{cccccc}
\psi_{11} & \cdots & \psi_{15} & \psi_{11'} & \psi_{12'} & \psi_{13'}\\
\psi_{21} & \cdots & \psi_{25} & \psi_{21'} & \psi_{22'} & \psi_{23'}\\
\psi_{31} & \cdots & \psi_{35} & \psi_{31'} & \psi_{32'} & \psi_{33'}\\
\psi_{41} & \cdots & \psi_{45} & \psi_{41'} & \psi_{42'} & \psi_{43'}\\
\psi_{51} & \cdots & \psi_{55} & \psi_{51'} & \psi_{52'} & \psi_{53'}\\
\frac{\partial}{\partial x} \psi_{11} &
\cdots &
\frac{\partial}{\partial x} \psi_{15} &
\frac{\partial}{\partial x} \psi_{11'} &
\frac{\partial}{\partial x} \psi_{12'} &
\frac{\partial}{\partial x} \psi_{13'}\\
\frac{\partial}{\partial x} \psi_{21} &
\cdots &
\frac{\partial}{\partial x} \psi_{25} &
\frac{\partial}{\partial x} \psi_{21'} &
\frac{\partial}{\partial x} \psi_{22'} &
\frac{\partial}{\partial x} \psi_{23'}\\
\frac{\partial}{\partial x} \psi_{31} &
\cdots &
\frac{\partial}{\partial x} \psi_{35} &
\frac{\partial}{\partial x} \psi_{31'} &
\frac{\partial}{\partial x} \psi_{32'} &
\frac{\partial}{\partial x} \psi_{33'}\\
\end{array} \right)
\left( \begin{array}{c}
\alpha_1\\
\alpha_2\\
\alpha_3\\
\alpha_4\\
\alpha_5\\
\alpha_{1'}\\
\alpha_{2'}\\
\alpha_{3'}
\end{array} \right).
\label{eq:image}
\end{equation}
Even if a node is located on a corner and thus subject to two different boundary
conditions at the same time [Fig.~\ref{fig:image}(b)], we can readily write down
a $10 \times 10$ collocation matrix which is non-singular.
The insertion of such a corner node is important in reducing numerical error,
because a well-known problem of the collocation method is that the result is the
most inaccurate near boundaries~\cite{Fasshauer2007}.

Note that Eq.~(\ref{eq:image}) takes care of both $T$ and its derivative on
an equal footing. Formally, we may consider images for \emph{every} sample node,
even if it belongs to $\Omega$, with setting their contributions to be trivially
zero. In this way, we merge Steps $4$ and $5$ in the previous section and
treat all the sample nodes with a single step.

\section{BENCHMARK TEST RESULTS}
\label{sec:bench}

\subsection{First Test: Boundary Value Problem}
As in Ref.~\citealp{Sarler2006}, we use the NAFEMS benchmark test
No.~10~\cite{Cameron1986}:
We consider a rectangular domain $\Omega = (0,L_x) \times (0,L_y)$ with $L_x =
0.6$m and $L_y = 1.0$m.
The material properties are specified by $k = 52$W~m$^{-1}$~\textdegree{C}$^{-1}$, $c = 460$J~kg$^{-1}$~\textdegree{C}$^{-1}$, and $\rho = 7850$kg~m$^{-3}$.
The temperature is fixed to $T_D = 100$\textdegree{C} of the lower boundary at
$y=0$. The left boundary at $x=0$ is thermally
insulated so that the proper choice is the Neumann boundary condition with
$\left. \partial T / \partial x \right|_{x=0} = 0$\textdegree{C}~m$^{-1}$.
On the other two boundaries, we have heat convection to $T_{\rm ref} =
0$\textdegree{C} with a convective heat transfer coefficient $h =
750$W~m$^{-2}$~\textdegree{C}$^{-1}$. It is expressed as a Robin boundary
condition [Eq.~(\ref{eq:robin})] with $R \equiv -h/k$.
Under these boundary conditions, the
analytic solution of the Laplace equation for $T(\mathbf{r})$ with $\mathbf{r}
\equiv (x,y)$ is given as
\begin{equation}
T_{\rm ana}(\mathbf{r})
= \sum_{n=1}^\infty \frac{ -2 T_D R \cos(\beta_n x) \{ \beta_n
\cos[\beta_n (L_y-y)] - R \sinh[\beta_n (L_y-y)]\}}{\cos(\beta_n L_x)
 [\beta_n \cosh(\beta_n L_y) - R \sinh(\beta_n L_y)] [L_x (R^2 + \beta_n^2) -
R]},
\label{eq:ana1}
\end{equation}
where $\beta_n$ is the $n$th positive root of the following equation
\begin{equation}
\beta \tan (\beta L_x) + R = 0.
\end{equation}

To check numerical performance, we are concerned with two quantities.
One is the maximum absolute deviation of
our numerical solution $T$ from the analytic solution $T_{\rm ana}$,
\begin{equation}
\Delta T_{\rm max} = \max \left| T_{\rm ana}(\mathbf{r}_n) - T(\mathbf{r}_n)
\right|,
\end{equation}
and the other is the average absolute deviation
\begin{equation}
\Delta T_{\rm avg} = \frac{1}{N} \sum_{n=1}^N \left| T_{\rm ana}(\mathbf{r}_n) -
T(\mathbf{r}_n) \right|,
\end{equation}
where $\mathbf{r}_n$ denotes the position of the node indexed as $n$.
On the other hand, we can try a quick check by measuring the temperature at a
reference point $\mathbf{r}_{\rm NAFEMS}$ with $x_{\rm NAFEMS}=0.6$m and $y_{\rm
NAFEMS}=0.2$m, whose analytic value is $T_{\rm NAFEMS} \approx 18.2538$
\textdegree{C} according to Eq.~(\ref{eq:ana1}).

\begin{figure}
\includegraphics[width=\textwidth]{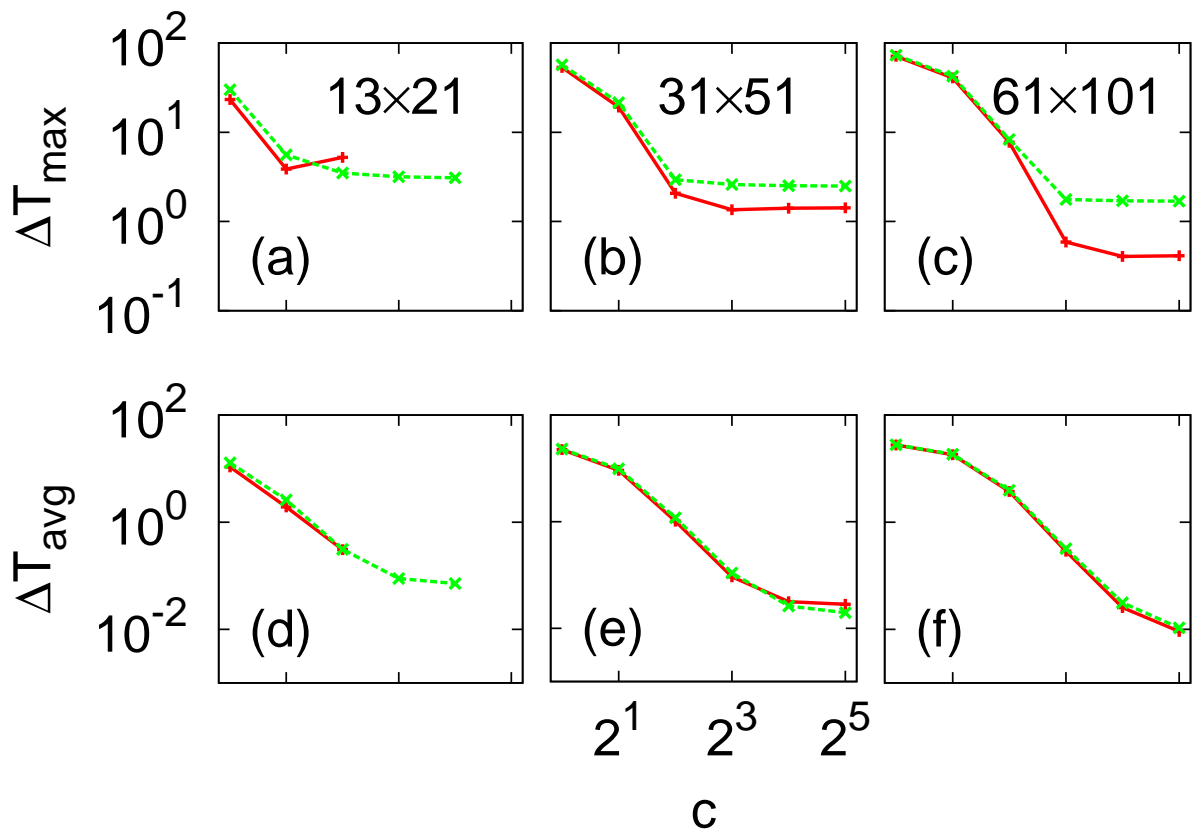}
\caption{The solid red lines represent the results in Tables~\ref{table:13x21},
\ref{table:31x51}, and \ref{table:61x101}, obtained with $K=5$.
The dotted green lines
are taken from Ref.~\citealp{Sarler2006} for comparison.
The upper and lower rows show
$\Delta T_{\rm max}$ and $\Delta T_{\rm avg}$, respectively, and the columns
mean different node arrangements from $13\times 21$ to $61 \times 101$.
The horizontal axis shows the values of $c = 1, 2, 4, \ldots, 32$.
A missing data point means that the solution diverges.
}
\label{fig:benchmark1}
\end{figure}

\begin{table}[tp]
\caption{First benchmark test result of the image method with domain size $K=5$
and $13 \times 21$ nodes. The last two columns show the position of the node
with the maximum absolute error $\Delta T_{\rm max}$. The method becomes
unstable for $c \ge 8$.}
\begin{tabular}{|c|c|c|c|c|}\hline
$c$ & $\Delta T_{\rm avg}$ [\textdegree C] & $\Delta T_{\rm max}$ [\textdegree
C] & $x_{\rm max}$ [m] & $y_{\rm max}$ [m]\\\hline
1  & 10.7990 & 23.2401 & 0.15 & 0.30\\\hline
2  & 1.9253 & 3.8687 & 0.15 & 0.35 \\\hline
4  & 0.3081 & 5.2266 & 0.60 & 0.05 \\\hline
\end{tabular}
\label{table:13x21}
\end{table}

\begin{table}[tp]
\caption{First benchmark test result of the image method with $31\times 51$
nodes.}
\begin{tabular}{|c|c|c|c|c|}\hline
$c$ & $\Delta T_{\rm avg}$ [\textdegree C] & $\Delta T_{\rm max}$ [\textdegree
C] & $x_{\rm max}$ [m] & $y_{\rm max}$ [m]\\\hline
1  & 22.5082 & 53.2592 &  0.10 & 0.20 \\\hline
2  & 9.1417  & 19.1447 &  0.06 & 0.32 \\\hline
4  & 1.0199  & 2.0731  &  0.06 & 0.36 \\\hline
8  & 0.0925  & 1.3545  &  0.60 & 0.02 \\\hline
16 & 0.0314  & 1.4061  &  0.60 & 0.02 \\\hline
32 & 0.0281  & 1.4229  &  0.60 & 0.02 \\\hline
\end{tabular}
\label{table:31x51}
\end{table}

\begin{table}[tp]
\caption{First benchmark test result of the image method with $61\times 101$
nodes.}
\begin{tabular}{|c|c|c|c|c|}\hline
$c$ & $\Delta T_{\rm avg}$ [\textdegree C] & $\Delta T_{\rm max}$ [\textdegree
C] & $x_{\rm max}$ [m] & $y_{\rm max}$ [m]\\\hline
1  & 27.6561 & 70.9615 & 0.09 & 0.13\\\hline
2  & 18.2390 & 40.5895 & 0.05 & 0.26\\\hline
4  & 3.7371  & 7.7168  & 0.03 & 0.36\\\hline
8  & 0.2880  & 0.5874  & 0.03 & 0.37\\\hline
16 & 0.0255  & 0.4048  & 0.60 & 0.01\\\hline
32 & 0.0092  & 0.4138  & 0.60 & 0.01\\\hline
\end{tabular}
\label{table:61x101}
\end{table}

\begin{table}[tp]
\caption{First benchmark test result at $\mathbf{r}_{\rm NAFEMS}$. The second
last column shows error from the analytic solution [Eq.~(\ref{eq:ana1})], and
the last column is taken from Ref.~\citealp{Sarler2006} for comparison.}
\begin{tabular}{|c|c|c|c|c|c|}\hline
Nodes & $K$ & $c$ & $T$[\textdegree C] & Error from $T_{\rm
NAFEMS}$[\textdegree C] &
Error in Ref.~\citealp{Sarler2006}[\textdegree C]\\\hline
$13 \times 21$  & $5$ & $4$  & 17.7508 & 0.5029 & 0.1075 \\\hline
$31 \times 51$  & $5$ & $32$ & 18.1846 & 0.0692 & 0.0317 \\\hline
$61 \times 101$ & $5$ & $32$ & 18.2375 & 0.0162 & 0.0056 \\\hline
\end{tabular}
\label{table:ref}
\end{table}

Figure~\ref{fig:benchmark1} summarizes our main results.
It is a graphical representation of the numerical data
tabulated in Tables~\ref{table:13x21}, \ref{table:31x51}, and
\ref{table:61x101}.
Note that the results are only for $K=5$ because our method is
unstable for $K=9$, whereas both the cases are available in
Ref.~\citealp{Sarler2006}. This may be an example of the trade-off
between accuracy and stability~\cite{Fasshauer2007}.
The figure shows that
our image method can significantly reduce the maximum absolute error $\Delta
T_{\rm max}$. For
example, for the node arrangement of $61 \times 101$, $\Delta T_{\rm max}$ is
reduced almost by a factor of $4$ compared with the results in
Ref.~\citealp{Sarler2006} [Fig.~\ref{fig:benchmark1}(c)].
It turns out essential to have corner nodes, such as the one indexed as
$3$ in Fig.~\ref{fig:image}(b), to reduce $\Delta T_{\rm max}$. Those corner
nodes can be properly handled by using image nodes, when they have
to satisfy more than one condition. Without the images,
the maximum absolute error would decrease rather slowly as the number of nodes
grows [see the dotted green lines in Figs.~\ref{fig:benchmark1}(a) to (c), which
depict the results in Ref.~\citealp{Sarler2006}].

Although the image method enhances accuracy in terms of this maximum absolute
error,
it increases numerical instability. For example, when we work with $13 \times
21$ nodes, our method give diverging results for $c = 8$
(Table~\ref{table:13x21}), whereas the results would converge without
the images~\cite{Sarler2006}.
In addition, we should note that the average error decreases only slightly
[Fig.~\ref{fig:benchmark1}(d) to (f)] and even increases sometimes
[see the rightmost points in Fig.~\ref{fig:benchmark1}(e)].
In Table~\ref{table:ref}, we check deviations from $T_{\rm NAFEMS}$ at
$\mathbf{r}_{\rm NAFEMS}$ for different node arrangements. It quickly decreases
as the number of nodes increases, but still greater than in
Ref.~\citealp{Sarler2006}.

\subsection{Second Test: Initial Value Problem}

Although we are primarily concerned about the boundary value problem, we have
also checked the initial value problem addressed in Ref.~\citealp{Sarler2006}
for completeness. We solve the diffusion equation
[Eq.~(\ref{eq:diff})] on a square domain with $L_x = L_y = 1.0$m. The
material properties take unit values, i.e., $\rho = 1$kg~m$^{-3}$,
$c=1$J~kg$^{-1}$~\textdegree{C}$^{-1}$, and
$k=1$W~m$^{-1}$~\textdegree{C}$^{-1}$. The boundary conditions are
also simplified so that the temperature on the right and upper boundaries is
fixed to $T_D = 0$\textdegree{C}, whereas the other two boundaries are of
the Neumann type with zero heat flux. If $T(\mathbf{r}) = 1$\textdegree{C} at
$t=0$, the analytic solution~\cite{Carslaw1959} is given as
\begin{equation}
T_{\rm ana}(\mathbf{r},t) = T_{\rm ana}(x,t) T_{\rm ana}(y,t),
\end{equation}
where
\begin{equation}
T_{\rm ana}(q,t) = \frac{4}{\pi} \sum_{n=0}^\infty \frac{(-1)^n}{2n+1}
\exp \left[ -\frac{k(2n+1)^2 \pi^2 t}{4\rho c L_q^2} \right] \cos \left[
\frac{(2n+1) \pi q}{2 L_q} \right],
\end{equation}
where $q$ means either $x$ or $y$.

\begin{figure}
\includegraphics[width=0.8\textwidth]{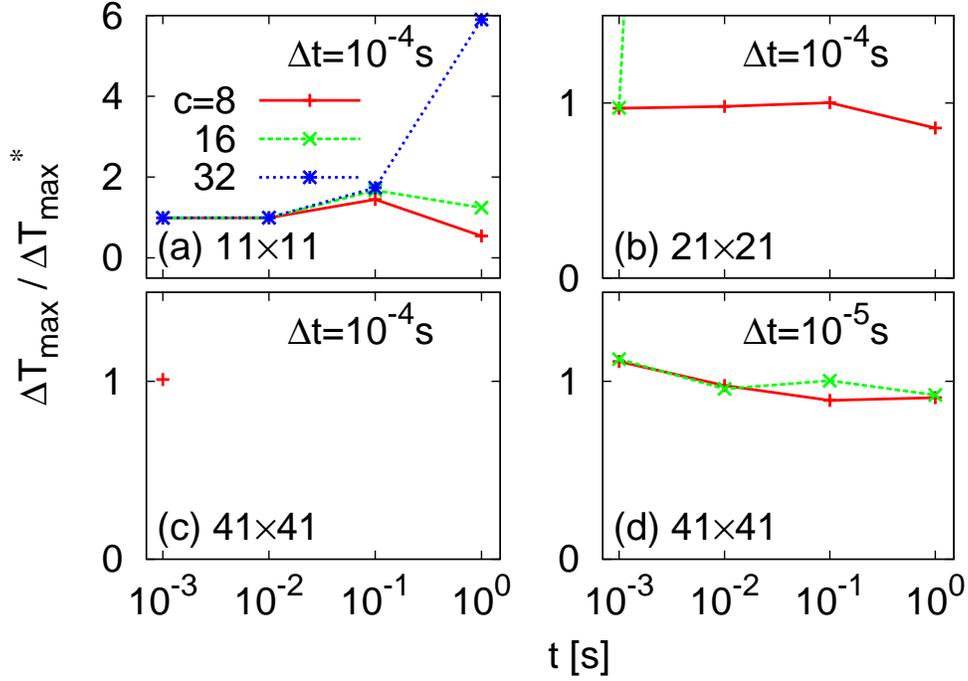}
\caption{The ratio of our maximum absolute error $\Delta T_{\rm max}$
with respect to that of Ref.~\citealp{Sarler2006}, denoted as $\Delta T_{\rm
max}^\ast$, in the second benchmark test. If the ratio is less than unity,
for example, it means that we have a more accurate result than in
Ref.~\citealp{Sarler2006}. All the results are obtained with $K=5$.
Each panel shows a
different combination of the node arrangement and the time step $\Delta t$ for
numerical integration. As in panel (a), a different color means a different
value of $c$, and a missing data point means that the solution diverges.}
\label{fig:benchmark2a}
\end{figure}

\begin{figure}
\includegraphics[width=0.8\textwidth]{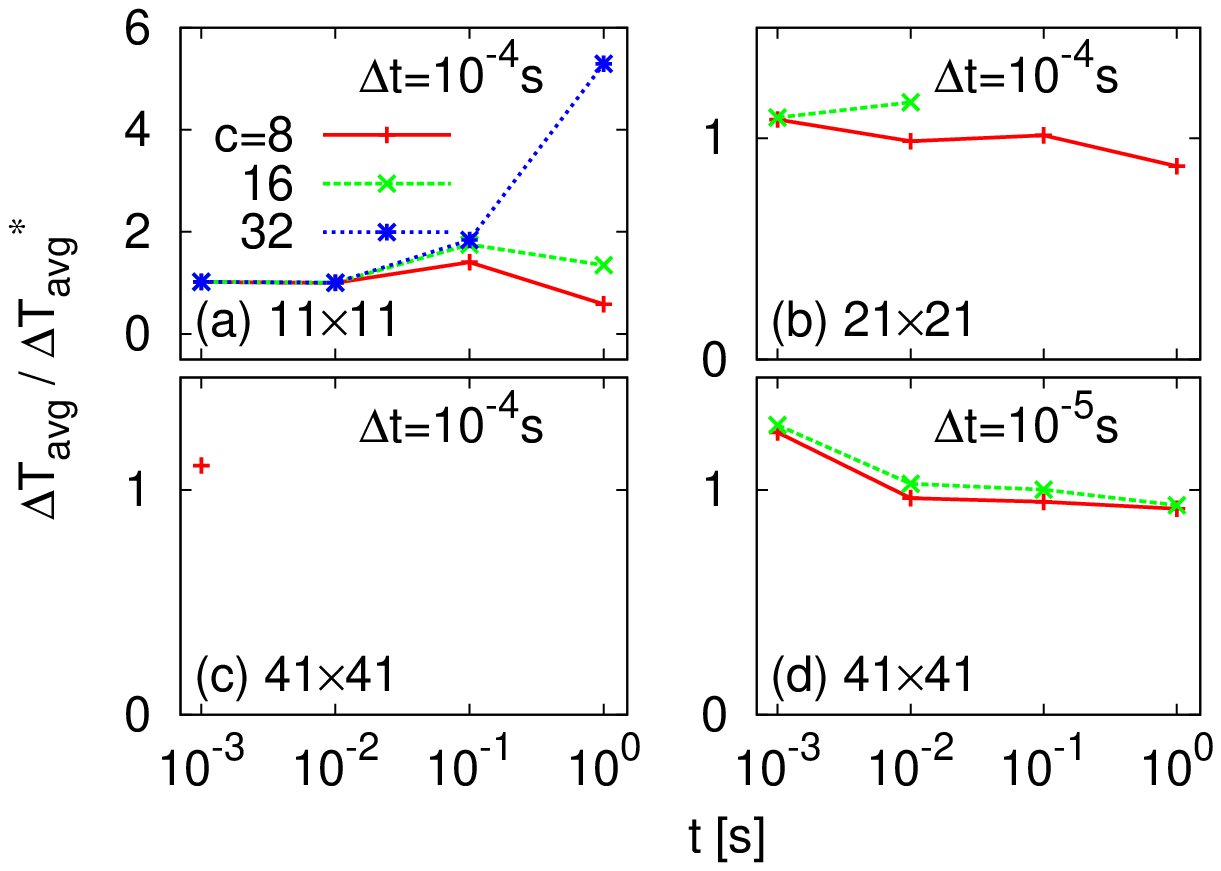}
\caption{The ratio of our average absolute error $\Delta T_{\rm avg}$
with respect to that of Ref.~\citealp{Sarler2006}, denoted as $\Delta T_{\rm
avg}^\ast$, in the second benchmark test. If the ratio is less than unity,
therefore, it means that we have a more accurate result than in
Ref.~\citealp{Sarler2006}. The other details are the same as
explained in the caption of Fig.~\ref{fig:benchmark2a}.}
\label{fig:benchmark2b}
\end{figure}

The results are tabulated in Tables~\ref{table:11x11} to \ref{table:41x41b}, and
their graphical representations are given in Figs.~\ref{fig:benchmark2a} and
\ref{fig:benchmark2b}. Overall, we get slightly better numerical accuracy than
in Ref.~\citealp{Sarler2006} when it comes to the largest number of sample nodes
and the smallest $\Delta t = 10^{-5}$s. The price is numerical
instability in that the result blows up with $c=32$ except for the smallest
number of sample nodes. This result is not very surprising, however, because the
method of images is meant to deal with more complicated boundary-value
problems.

\begin{table}[hp]
\caption{Second benchmark test result of the image method with $\Delta t =
10^{-4}s$ and $11 \times 11$ nodes. The size of $\omega$ is set to be $K=5$
for every $l$.}
\begin{tabular}{|c|c|c|c|c|c|}\hline
$t$ [s] & $c$ & $\Delta T_{\rm avg}$ [\textdegree C] & $\Delta T_{\rm max}$
[\textdegree C] & $x_{\rm max}$ [m] & $y_{\rm max}$ [m]\\\hline
$10^{-3}$ &  8 & 1.206e-02 & 1.245e-01 & 0.900 & 0.900\\\hline
$10^{-3}$ & 16 & 1.204e-02 & 1.243e-01 & 0.900 & 0.900\\\hline
$10^{-3}$ & 32 & 1.204e-02 & 1.243e-01 & 0.900 & 0.900\\\hline
$10^{-2}$ &  8 & 4.864e-03 & 2.265e-02 & 0.700 & 0.700\\\hline
$10^{-2}$ & 16 & 4.787e-03 & 2.231e-02 & 0.700 & 0.700\\\hline
$10^{-2}$ & 32 & 4.769e-03 & 2.222e-02 & 0.700 & 0.700\\\hline
$10^{-1}$ &  8 & 1.743e-03 & 5.015e-03 & 0.000 & 0.100\\\hline
$10^{-1}$ & 16 & 1.250e-03 & 4.330e-03 & 0.000 & 0.100\\\hline
$10^{-1}$ & 32 & 1.148e-03 & 4.168e-03 & 0.000 & 0.100\\\hline
$10^{0} $ &  8 & 4.071e-05 & 9.704e-05 & 0.000 & 0.000\\\hline
$10^{0} $ & 16 & 1.222e-05 & 3.498e-05 & 0.100 & 0.000\\\hline
$10^{0} $ & 32 & 2.477e-05 & 6.604e-05 & 0.100 & 0.000\\\hline
\end{tabular}
\label{table:11x11}
\end{table}

\begin{table}[hp]
\caption{Second benchmark test result of the image method with $\Delta t =
10^{-4}s$ and $21 \times 21$ nodes. The size of $\omega$ is set to be $K=5$
for every $l$. For $c=16$ and $32$, the solution diverges as time goes by.}
\begin{tabular}{|c|c|c|c|c|c|}\hline
$t$ [s] & $c$ & $\Delta T_{\rm avg}$ [\textdegree C] & $\Delta T_{\rm max}$
[\textdegree C] & $x_{\rm max}$ [m] & $y_{\rm max}$ [m]\\\hline
$10^{-3}$ &  8 & 4.984e-03 & 4.283e-02 & 0.900 & 0.900\\\hline
$10^{-3}$ & 16 & 4.949e-03 & 4.257e-02 & 0.900 & 0.900\\\hline
$10^{-3}$ & 32 & 5.254e-03 & 1.356e-01 & 0.050 & 0.000\\\hline
$10^{-2}$ &  8 & 1.553e-03 & 6.983e-03 & 0.750 & 0.750\\\hline
$10^{-2}$ & 16 & 1.545e-03 & 7.971e-02 & 0.050 & 0.000\\\hline
$10^{-1}$ &  8 & 1.258e-03 & 2.390e-03 & 0.250 & 0.100\\\hline
$10^{0} $ &  8 & 9.355e-05 & 2.310e-04 & 0.000 & 0.000\\\hline
\end{tabular}
\label{table:21x21}
\end{table}

\begin{table}[hp]
\caption{Second benchmark test result of the image method with $\Delta t =
10^{-4}s$ and $41 \times 41$ nodes. The size of $\omega$ is set to be $K=5$
for every $l$. The result diverges for $c=16$ and $32$, and it is the case even
for $c=8$ when $t \gtrsim 10^{-2}$.}
\begin{tabular}{|c|c|c|c|c|c|}\hline
$t$ [s] & $c$ & $\Delta T_{\rm avg}$ [\textdegree C] & $\Delta T_{\rm max}$
[\textdegree C] & $x_{\rm max}$ [m] & $y_{\rm max}$ [m]\\\hline
$10^{-3}$ &  8 & 1.993e-03 & 2.428e-02 & 0.950 & 0.950\\\hline
\end{tabular}
\label{table:41x41a}
\end{table}

\begin{table}[hp]
\caption{Second benchmark test result of the image method with $\Delta t =
10^{-5}s$ and $41 \times 41$ nodes. The size of $\omega$ is set to be $K=5$
for every $l$. We see diverging results for $c=32$.}
\begin{tabular}{|c|c|c|c|c|c|}\hline
$t$ [s] & $c$ & $\Delta T_{\rm avg}$ [\textdegree C] & $\Delta T_{\rm max}$
[\textdegree C] & $x_{\rm max}$ [m] & $y_{\rm max}$ [m]\\\hline
$10^{-3}$ &  8 & 1.556e-03 & 1.745e-02 & 0.925 & 0.925\\\hline
$10^{-3}$ & 16 & 1.519e-03 & 1.688e-02 & 0.925 & 0.925\\\hline
$10^{-2}$ &  8 & 7.795e-04 & 2.577e-03 & 0.775 & 0.775\\\hline
$10^{-2}$ & 16 & 3.932e-04 & 1.696e-03 & 0.750 & 0.750\\\hline
$10^{-1}$ &  8 & 1.865e-03 & 3.448e-03 & 0.225 & 0.100\\\hline
$10^{-1}$ & 16 & 3.160e-04 & 6.349e-04 & 0.125 & 0.025\\\hline
$10^{0} $ &  8 & 1.892e-04 & 4.676e-04 & 0.000 & 0.000\\\hline
$10^{0} $ & 16 & 2.441e-05 & 6.034e-05 & 0.000 & 0.000\\\hline
\end{tabular}
\label{table:41x41b}
\end{table}

\section{DISCUSSION AND SUMMARY}
\label{sec:conclusion}

In summary, we have modified the local RBF collocation method by adding an
additional set of nodes in the same spirit of PDE collocation on the boundary in
Ref.~\citealp{Fedoseyev2002}. This method
makes it possible to take into account every piece of available
information on the boundaries. That is, our collocation matrix can describe
both the functional value $T$ as well as its spatial derivatives on every
boundary node even if the node is subject to two or more boundary conditions.
This small modification is able to reduce the maximum error $\Delta T_{\rm max}$
relative to the analytic solution almost by a factor of $4$ in the first
benchmark test for a boundary value problem [see Fig.~\ref{fig:benchmark1}(c)].
It should be noted that the collocation at the boundaries makes the numerical
integration more unstable. It is therefore desirable to use more sample nodes
and smaller time steps for convergence, and one could think of implementing an
implicit scheme such as the Crank-Nicholson method rather than our simple Euler
scheme. We do not pursue this direction because our purpose is to make a direct
comparison with Ref.~\citealp{Sarler2006}.
Overall, if a boundary value problem is given with Robin boundary conditions, we
can recommend including collocation at the boundaries: Combined with the local
RBF collocation method, the additional amount of effort is small whereas the
reduction of the maximum error is significant, as long as the result is
convergent with a sufficiently large number of sample nodes.

\begin{acknowledgments}
This work was supported by a research grant of Pukyong National University
(2015).
\end{acknowledgments}


%

\end{document}